\documentclass[english]{achemso}
\usepackage[LGR,T1]{fontenc}
\usepackage[latin9]{inputenc}
\usepackage{geometry}
\usepackage{array}
\usepackage{booktabs}
\usepackage{textcomp}
\usepackage{pdfpages}
\usepackage{multirow}
\usepackage{amsmath}
\usepackage{amssymb}
\usepackage{graphicx}
\usepackage[numbers]{natbib}
\PassOptionsToPackage{normalem}{ulem}
\usepackage{ulem}

\makeatletter

\title{Enhanced hydrogen evolution reaction activity of nitrogen deficient
hg-C$_{3}$N$_{4}$ quantum dot}

\author{Khushboo Dange}

\email{khushboodange@gmail.com}

\affiliation{Department of Physics, Indian Institute of Technology Bombay, Powai,
Mumbai 400076, India}

\author{Vaishali Roondhe}

\email{oshivaishali@gmail.com}

\affiliation{Department of Physics, Indian Institute of Technology Bombay, Powai,
Mumbai 400076, India}

\author{Alok Shukla}

\email{shukla@iitb.ac.in}

\affiliation{Department of Physics, Indian Institute of Technology Bombay, Powai,
Mumbai 400076, India}

\DeclareRobustCommand{\greektext}{%
  \fontencoding{LGR}\selectfont\def\encodingdefault{LGR}}
\DeclareRobustCommand{\textgreek}[1]{\leavevmode{\greektext #1}}
\ProvideTextCommand{\~}{LGR}[1]{\char126#1}

\newcommand{\lyxmathsym}[1]{\ifmmode\begingroup\def\b@ld{bold}
  \text{\ifx\math@version\b@ld\bfseries\fi#1}\endgroup\else#1\fi}

\providecommand{\tabularnewline}{\\}

\mciteErrorOnUnknownfalse

\makeatother

\usepackage{babel}
\begin{document}
\begin{abstract}
The present study investigates the catalytic performance of a hg-C$_{3}$N$_{4}$
quantum dot aimed at enhancing electrochemical water splitting, using
the first-principles density functional theory. The size of the considered
quantum dot lies within the range reported experimentally (2nm --
4nm) {[}Zhou \textit{et al}. ACS Nano\textbf{ 9}, 12480 (2015){]}.
The nitrogen vacancies are created in the considered hg-C$_{3}$N$_{4}$
structure to simulate the realistic scenario, as the presence of nitrogen
and carbon defects are reported in the synthesized hg-C$_{3}$N$_{4}$
quantum dots. First, the structural and vibrational properties are
computed to ensure the stability of the nitrogen-deficient hg-C$_{3}$N$_{4}$
quantum dots, and subsequently, their electronic and hydrogen evolution
reaction (HER) properties are investigated. The calculated HER parameters,
i.e., adsorption energies, Gibbs free energies, and overpotentials
demonstrate that the considered hg-C$_{3}$N$_{4}$ quantum dot with
nitrogen vacancies can be used as a moderately effective electrocatalyst
for HER performance. We also considered the quantum dot to be dissolved
in water and ethanol, and find that the overpotential gets drastically
reduced to 16 mV for the alcohol dissolved QD, while some significant
reduction is seen for the aqueous solution also. As a result, this
study suggests that the nitrogen-deficient hg-C$_{3}$N$_{4}$ quantum
dots dissolved in ethanol are excellent candidates for catalysis aimed
at sustainable hydrogen production via electrochemical water splitting.
\end{abstract}

\section{Introduction}

The increasing energy crisis and accompanying environmental issues
forced scientists to develop alternative, non-conventional sources
of energy. The use of environment-friendly energy resources will not
only allow us to overcome the pressing issue of the depletion of traditional
energy resources, but it will also contribute to the development of
a pollution-free environment. Hydrogen is one such renewable and pure
form of energy resource. Its high gravimetric energy density ($120$
$kJg^{-1}$) \citep{MOLLER2017}, recyclability, and environmental
friendliness make it a potential candidate for future fuel \citep{WALLACE1983,ISHAQ2022}.
The major challenge faced by the scientists is to make hydrogen freely
available on earth in appreciable or exploitable concentrations. More
and more attention has been paid over the last few years to the development
of efficient and reliable technologies for the production of clean
and sustainable hydrogen. The electrochemical water splitting has
been considered as an efficient technique to produce hydrogen \citep{CHEN2021},
as it does not lead to the emission of hazardous gases, instead, it
restores oxygen to the environment. This is the main driving force
which encourages scientists to investigate this particular technology.
However, a large overpotential is needed to overcome the activation
barrier for the two half reactions, i.e., the hydrogen evolution reaction
(HER) and the oxygen evolution reaction (OER) \citep{Li2020,CHENG2024},
which makes electrochemical water splitting an expensive technique.
As a consequence, there is a tremendous need to reduce the overpotential
required for HER, and therefore, make it cost effective. The reduction
in overpotential can be achieved using an active catalyst, and thus
opens a way for efficient hydrogen production \citep{CHEN2021}. 

Pt-based materials are been considered to be the most effective catalysts
for HER \citep{Cheng2016,Huang2013}. However, the scarcity and high
cost of noble metals severely limit their use as catalysts. Therefore,
research in the field of active catalysts that are inexpensive, earth-abundant,
and easy to synthesize is growing at a fast pace, as it will lead
to economically feasible HER \citep{Hunter2016,C2EE03250C,LI2024}.
Nanomaterials can be quite efficient as catalysts because of their
significant advantages, such as low fabrication cost, high specific
surface area, and large number of active sites compared to their bulk
form. So far, various inexpensive low-dimensional materials are being
studied as electrocatalysts of industrial importance for HER \citep{wang2020,Chen2017,Babu2015}.
These nanocatalysts include nonmetals like carbon-based materials
such as graphene and its hybrids \citep{li2017,Qin2022,SHARMA2022}
and non-noble metals like transition metal-based materials such as
chalcogenides \citep{thomas2007}, nitrides \citep{popczun2013},
carbides \citep{yu2020}, sulfides \citep{kin2020}, and phosphides
\citep{limang2022}. Although pristine carbon materials are poor electrocatalysts
or electrochemically inert, their inertness can be reduced by doping
nonmetallic elements \citep{Zhou2015,Duan2015} or transition metals
\citep{Zhou2015}, thereby improving their catalytic performance.
Nitrogen-doped carbon materials gained a lot of research attention
due to their various applications in the fields of photocatalysis
\citep{NASIR2020}, oxygen reduction reaction (ORR) \citep{liang2013n},
solar power generation \citep{li2020im}, etc. The difference in electronegativity
and size between the nitrogen and carbon atoms can polarize the carbon
atoms adjacent to the doped nitrogen atoms and lead to an increase
in the number of active sites. The large number of active sites are
advantageous to facilitate catalytic activities for HER, OER, and
ORR \citep{Garlyyev2019,zhang2016,Zheng2014,Suen2017,Wang2023,parman2021}. 

One such nitrogen-doped carbon compound is graphitic carbon-nitride
(g-C$_{3}$N$_{4}$), a metal-free, cost-effective, and earth-abundant
material. Owing to its $\pi$-conjugated structure, g-C$_{3}$N$_{4}$
is the most stable allotrope of carbon-nitride structures \citep{zhang2001}.
It has been confirmed by thermogravimetric analysis that \sout{\mbox{$hg-C_{3}N_{4}$}
}heptazine based g-C$_{3}$N$_{4}$ (hg-C$_{3}$N$_{4}$) is thermally
stable in air up to 600 $^{\text{\textgreek{o}}}C$, due to the presence
of aromatic $C-N$ heterocycles \citep{Ong2016}. A single layer of
g-C$_{3}$N$_{4}$ consists of two types of tectonic units, i.e.,
s-triazine and tri-s-triazine (or heptazine) \citep{KROKE2004}. However,
the tri-s-triazine unit has proven to be more stable \citep{Zhu2018}.
2D g-C$_{3}$N$_{4}$ and g-C$_{3}$N$_{4}$quantum dots (QD) with
tunable extraordinary geometric, electronic, and optical properties
and particularly large surface areas provide relatively more active
sites for H-adsorption. Zheng \textit{et al.} \citep{yao2014} reported
that g-C$_{3}$N$_{4}$ coupled with nitrogen-doped graphene results
in the formation of a metal-free hybrid catalyst and shows high electrocatalytic
performance for HER with an overpotential of 240 mV. Geng \textit{et
al.} \citep{GENG2019} designed MoS$_{2}$/S-doped g-C$_{3}$N$_{4}$
layered heterostructure and reported a low overpotential of 173 mV.
However, till date, hg-C$_{3}$N$_{4}$ quantum dots have rarely been
investigated as electrocatalyst for HER. g-C$_{3}$N$_{4}$ QDs (both
triazine and heptazine based) exhibit interesting geometric, electronic
and other properties, as a result of quantum confinement \citep{ZHAI2018,GHASHGHAEE,Ghashghaee2020}.
Research in the field of \sout{heptazine based g-C\mbox{$_{3}$}N\mbox{$_{4}$}
(hg-C\mbox{$_{3}$}N\mbox{$_{4}$})} hg-C$_{3}$N$_{4}$ QDs is growing
significantly on both the experimental and computational fronts \citep{ZHAI2018,wang2014,Zhou2015,zhou2013},
as they are more reactive due to the presence of large number of lone
pairs \citep{Ghashghaee2020,wang2012}. Furthermore, studies based
on the photocatalytic activity of hg-C$_{3}$N$_{4}$ QDs are also
present in the literature \citep{NASIR2020,YAN2017}, however, their
electrocatalytic properties have rarely been studied. Yang \textit{et
al.} \citep{YANG2022} recently fabricated ultrafine g-C$_{3}$N$_{4}$
QD and reported a low overpotential (208 mV), indicating good catalytic
performance for HER. In their work, the presence of a large number
of carbon and nitrogen defects in g-C$_{3}$N$_{4}$ QD is also confirmed
by the UV-vis spectrum. Apart from this, the literature does not contain
any further experimental or theoretical findings on the electrocatalytic
performance of g-C$_{3}$N$_{4}$ QDs. The lack of understanding of
the electrocatalytic performance of g-C$_{3}$N$_{4}$ QDs drives
our attention towards this. Since the experimental realization of
novel catalysts is difficult and time-consuming, computational investigations
using density functional theory would be beneficial in this regard.
As a consequence, in this work, we have calculated the electronic
and electrocatalytic properties of hg-C$_{3}$N$_{4}$ QDs, taking
into consideration the presence of nitrogen vacancies in the structure.
The existence of defects leads to lattice disorder and thereby increases
the active sites \citep{TIAN2017}, which is desirable for effective
HER performance. Our theoretical findings predicted somewhat unfavorable
catalytic performance of nitrogen-deficient hg-C$_{3}$N$_{4}$ QD
with moderate overpotential, which gets lowered drastically (significantly)
by considering it to be dissolved in ethanol (water). We believe that
our work will be the first theoretical investigation on the application
of hg-C$_{3}$N$_{4}$ QDs for HER. 

\section{Computational Methodology}

All the calculations, including structural, electronic, and HER-related
properties, were performed using the first-principles density functional
theory (DFT) \citep{hohenberg1964,kohn1965}, and Gaussian16 program
package \citep{gaussian16} was used for the purpose. The Gaussian-type
double zeta basis set 6-31G along with the two polarization functions
(d, p) was employed in all the calculations. This basis set takes
into account one Slater type orbital (STO) with six primitive Gaussians
for each inner shell and divides each valence shell into an inner
and outer part using three and one primitive Gaussians, respectively.
The convergence criteria for the energy and the minimum gradient forces
between two consecutive atoms were set to $10^{-8}$ Hartree and 0.000450
Hartree/Bohr, respectively. Since the present work deals with the
systems involving adsorption of atoms (H-atom), van der Waal interactions
were also taken into account. The wB97XD \citep{Chai2008} hybrid
functional was chosen for the purpose as it includes Grimme\textquoteright s
D2 dispersion corrections. This functional contains $22$ \% Hartree-Fock
(HF) exchange part at the short range, while it is entirely (i.e.
100\%) HF exchange in the long-range region. The initial formation
and visualization of the structures and corresponding molecular orbitals
were done using GaussView6 \citep{denn2009} software. Multiwfn software
\citep{multiwfn} was employed to generate both the total and projected
density of states. Finally, the IEFPCM \citep{IEFPCM1981} model has
been used to study the impact of solvent.

The working geometry is obtained by creating nitrogen vacancies in
the considered structure of hg-C$_{3}$N$_{4}$ QD. The formation
energy, $E_{form}$ of this optimized defective system named ``NV-hg-C$_{3}$N$_{4}$''
QD is calculated using the expression:

\begin{equation}
E_{form}=\frac{1}{N}(E_{defect}-E_{pristine}+\epsilon\mathit{\Delta}n)\label{eq:1}
\end{equation}

where $E_{pristine}$ and $E_{defect}$ are the ground state energies
of the system without defects and with defects (nitrogen-vacancies),
respectively. N represents the total number of atoms present in the
pristine structure (without defects), while $\epsilon$ and $\mathit{\Delta}n$
denote the energy of an isolated nitrogen atom and the number of nitrogen
atoms removed, respectively. To ensure the thermodynamic stability
of the defective system and all the considered configurations obtained
by adsorption of an H-atom at different sites, their cohesive energies
per atom are calculated using the expression

\begin{equation}
E_{cohesive}=\frac{1}{N}(E_{Total}-\underset{i}{\sum}\epsilon_{i}n_{i})\label{eq:2}
\end{equation}

where $N=\sum n_{i}$, and $n_{i}$ denotes the number of atoms of
particular type with their energies $\epsilon_{i}$, and $E_{Total}$
is the energy of the system. To ensure the dynamical stability of
the systems into consideration, we have also computed their vibrational
properties. 

Afterwards, the hydrogen adsorption energy ($\Delta E_{ads}$), which
is a crucial descriptor for HER, is calculated for the H-adsorbed
configurations as:

\begin{equation}
\Delta E_{ads}=E_{Total}-(E_{defect}+\epsilon_{H})\label{eq:3}
\end{equation}

where \sout{\mbox{$E_{total}$}} $E_{Total}$ here denotes total
energy of the NV-hg-C$_{3}$N$_{4}$ QD with adsorbed hydrogen, and
$\epsilon_{H}$ represents the energy of an isolated H-atom. The catalytic
performance of a material can be predicted theoretically from the
overpotential ($\eta$) in which Gibbs free energy ($\Delta G$) plays
a significant role. $\Delta G$ of an H-atom in the adsorbed state
can be computed using the equation \citep{Norskov2005}
\begin{equation}
\Delta G=\Delta E_{ads}+\Delta E_{ZPE}+T\Delta S\label{eq:Gibbs}
\end{equation}

where $\Delta S$ ($\Delta E_{ZPE}$) denotes the entropy (zero point
energy (ZPE)) difference between the adsorbed state of the system
and gaseous phase at temperature, T = 300 K. $\Delta E_{ZPE}$, which
ranges from 0.01 eV to 0.04 eV for the hydrogen molecule, can be given
as \citep{Tsai2014}
\begin{equation}
\Delta E_{ZPE}=\Delta E_{ZPE}^{nH}-\Delta E_{ZPE}^{(n-1)H}-\frac{1}{2}\Delta E_{ZPE}^{H_{2}}
\end{equation}

where $\Delta E_{ZPE}^{nH}$ and $\Delta E_{ZPE}^{(n-1)H}$ represent
ZPE corrections of the total energy of the system with $n$ and $n-1$
hydrogen atoms adsorbed on the catalyst, respectively, and $\Delta E_{ZPE}^{H_{2}}$
denotes ZPE of the gaseous phase of $H_{2}$. Further, $\Delta S$
in Eq. \ref{eq:Gibbs} can be approximated as 
\begin{equation}
\Delta S=(S_{nH}-S_{(n-1)H}-\frac{1}{2}S_{H_{2}})\approx\frac{1}{2}S_{H_{2}}^{0}
\end{equation}

where $S_{nH}$ and $S_{(n-1)H}$ signify the entropy of the system
with $n$ and $n-1$ adsorbed hydrogen atoms, respectively, and $\frac{1}{2}S_{H_{2}}$
represents the entropy of the gaseous phase of $H_{2}$. Finally,
$\Delta S$ of Eq. \ref{eq:Gibbs} is approximated as $\frac{1}{2}S_{H_{2}}^{0}$,
where $S_{H_{2}}^{0}$ denotes the vibrational entropy of the hydrogen
molecule leading to the value $TS_{H_{2}}^{0}$ = 0.41 eV. Further,
on taking $\Delta E_{ZPE}=0.04$ eV, we obtain the final expression
\begin{equation}
\Delta G=\Delta E_{ads}+0.24\label{eq:4}
\end{equation}

and then, $\eta$ can be calculated as
\begin{equation}
\eta=-\frac{\left|\Delta G\right|}{e}\label{eq:5}
\end{equation}

A positive overpotential is necessary for all anodic reactions, whereas
a negative overpotential is required for all cathodic reactions, and
thus for HER. However, for the sake of brevity, the absolute values
of the overpotential are referred throughout this article. 

\section{Results and Discussion}

This section is organized in a manner such that the structural and
vibrational properties are considered first, including the practical
feasibility of the structures. Following that, the energies corresponding
to the adsorption of an H-atom at different sites and the electronic
properties for the HER favorable cases (according to $\Delta E_{ads}$)
are discussed. Afterwards, we emphasize the HER-related properties
of the considered system, which is the primary objective of this research.

\subsection{Structural and Vibrational Properties}

\subsubsection{Optimized Pristine Geometry }

\begin{figure}[H]
\begin{centering}
\includegraphics[scale=0.7]{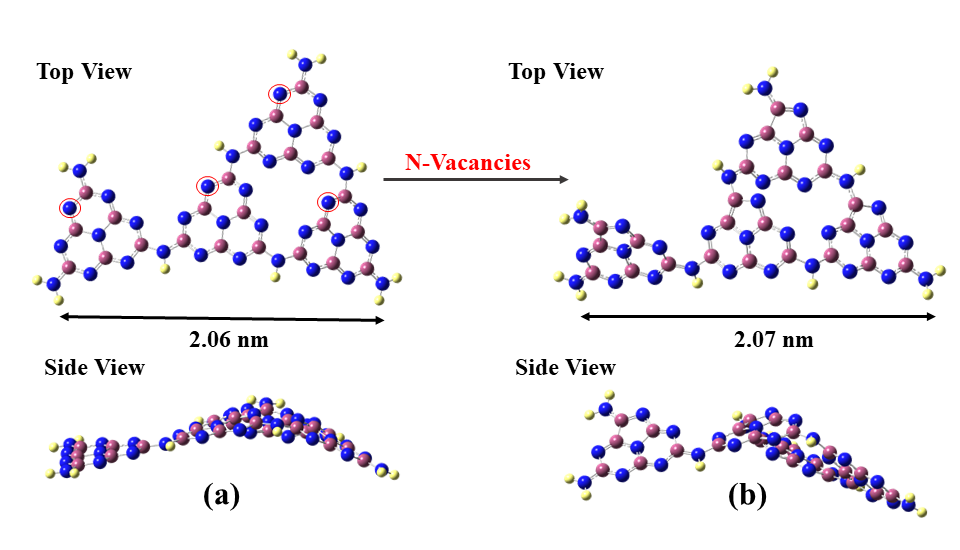}
\par\end{centering}
\caption{\label{fig:1}Optimized structures of (a) hg-C$_{3}$N$_{4}$ and
(b) NV-hg-C$_{3}$N$_{4}$ QDs. Here blue, pink, and yellow spheres
represent the nitrogen, carbon and hydrogen atoms, respectively, while
the red circles signify removal of that particular nitrogen atom.}
\end{figure}

\begin{figure}[H]
\begin{centering}
\includegraphics[scale=0.7]{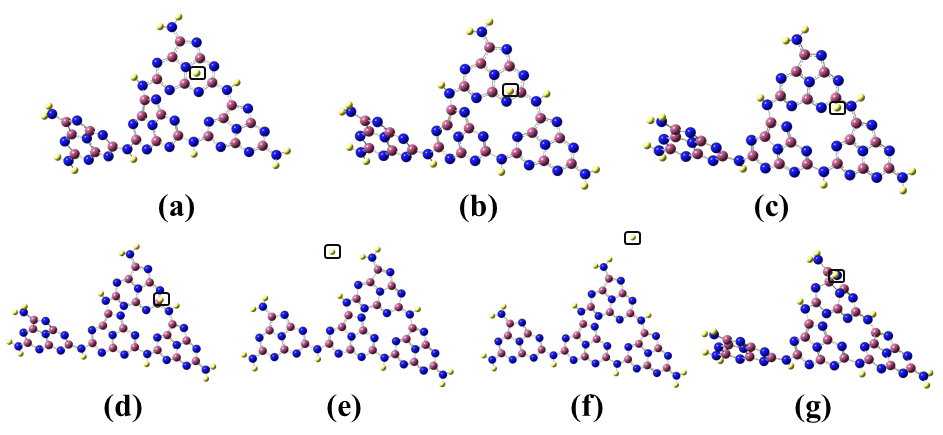}
\par\end{centering}
\caption{\label{fig:2}Optimized structures of H-adsorbed configurations of
NV-hg-C$_{3}$N$_{4}$ QD at the sites: (a) $top-C$, (b) $top-N$,
(c) $hole$, (d) $edge-C$, (e) $edge-N$, (f) $edge-H$, and (g)
$penta-C$. The final position occupied by adsorbed H-atom in each
of the cases is shown by a rectangular region. Here blue, pink, and
yellow spheres represent the nitrogen, carbon and hydrogen atoms,
respectively.}
\end{figure}

The considered hg-C$_{3}$N$_{4}$ QD structure comprises of four
heptazine units, as shown in Fig. \ref{fig:1}(a). Its geometry was
optimized in our previous work \citep{Dange2024} which confirmed
that the structure gets buckled due to the presence of lone pair on
nitrogen atoms. In this work, in order to consider the real-life situation,
structures with nitrogen vacancies are also considered. The nitrogen
sites have three different possibilities (sites 1--3) in the single
heptazine unit, as is clear from Fig.S1 of the supporting information
(SI). Thus, the vacancy formation at each site is considered to decide
which nitrogen site would be energetically favorable. The defect formation
energies, calculated using Eq.\ref{eq:1} for the sites 1--3 are
$0.19$ eV, $0.08$ eV, and $0.29$ eV, respectively. This implies
that site-2 is optimal for vacancy formation. Consequently, it is
decided to introduce N-vacancies at site-2 of each heptazine unit
of the considered hg-C$_{3}$N$_{4}$ QD, leading four vacancies in
all. The resulting optimized geometry is shown in Fig.\ref{fig:1}(b).
This N-vacated hg-C$_{3}$N$_{4}$ (NV-hg-C$_{3}$N$_{4}$) QD is
considered in the present study to investigate its electrocatalytic
performance for HER along with its electronic properties. The optimized
geometry of the defective structure exhibits more buckling as compared
to the pristine one. Also, each hexagon from which a nitrogen atom
is removed gets converted into a pentagon. The formation of a pentagon
due to the removal of an atom from a hexagon of C atoms has already
been reported in the literature, which validates our result \citep{penta-hepta2008}.
The larger extent of the buckling of the $sp^{2}$-coordinated basal
plane is due to the incorporation of pentagons in the structure. The
size of the considered NV-hg-C$_{3}$N$_{4}$ QD is $2.07$ nm, and
this lies within the experimentally observed range as far as the size
of g-C$_{3}$N$_{4}$ QDs is concerned \citep{Zhixin2015,juan2013}.
The measured average $C-N$ and $C-C$ (in pentagon) bond lengths
are $1.36$ Å and $1.40$ Å, respectively, and the bond angles $\angle CNC$
and $\angle NCN$ are observed to be $117\lyxmathsym{º}$ and $120\text{º}$,
respectively. The H-atom to be adsorbed is placed on top of different
sites of the NV-hg-C$_{3}$N$_{4}$ QD as shown in Fig.S2 of the SI.
The considered sites include: (a) C/N/H atoms located at the edge
(\emph{edge-C/N}/\emph{H}), (b) interior atoms (\emph{top-C/N}), (c)
center of the hole formed by the three inter-connected heptazine units
($hole$), and (d) a carbon atom of the pentagons ($penta-C$). After
optimization, the final positions of the adsorbed H-atom in all the
cases are illustrated in Fig.\ref{fig:2}. When placed at the $top-C$
and $top-N$ sites, the ground state configuration is achieved when
the adsorbed H-atom gets shifted at some position inside the corresponding
hexagonal ring (Fig.\ref{fig:2}(a) and (b)). When placed at the center
of a hole, the H-atom gets shifted to the boundary of the hole (Fig.\ref{fig:2}(c)).
In case of the $edge-C$ site, the H-atom occupies a position over
the top of an adjacent carbon atom after optimization (Fig.\ref{fig:2}(d)).
However, for the $edge-N$ ($edge-H$) site, the adsorbed H-atom gets
repelled and attains a position at significantly larger distance of
4.2 Å (3.4 Å) from the structure, as depicted in Fig.\ref{fig:2}(e)
(Fig.\ref{fig:2}(f)), suggesting no adsorption. In case of the $penta-C$
configuration, the H-atom forms a bond with the carbon atom on which
it is placed, indicating strong adsorption (Fig.\ref{fig:2}(g)).
The adsorption energies will be calculated later to confirm this findings.
Initially, the adsorption and Gibbs free energies were calculated
for the pristine hg-C$_{3}$N$_{4}$ QD but the obtained values were
quite large, and therefore, not suitable for HER activity (see SI).
Therefore, we considered the N-vacated structure for further study
in this work.

\subsubsection{Practical Feasibility}

Although defective hg-C$_{3}$N$_{4}$ QDs have already been synthesized
\citep{YANG2022}, still the stability of our designed NV-hg-C$_{3}$N$_{4}$
QD is an important factor to be determined. To analyze the thermodynamic
stability of the NV-hg-C$_{3}$N$_{4}$ QD under consideration, we
have calculated both the defect formation energy ($E_{form}$) and
cohesive energy ($E_{cohesive}^{(d)}$) using Eq.\ref{eq:1} and Eq.\ref{eq:2},
respectively, and the obtained values are $0.72$ eV/atom and $-7.26$
eV/atom, respectively. The negative $E_{cohesive}^{(d)}$ and a small
positive value of $E_{form}$ are in favor of the thermodynamic stability
and practical feasibility of the considered NV-hg-C$_{3}$N$_{4}$
structure. The relative cohesive energy for this defective system,
$\Delta E_{cohesive}=E_{cohesive}^{(d)}-E_{cohesive}^{(p)}$ is also
calculated with respect to the cohesive energy of the pristine structure
($E_{cohesive}^{(p)}=-7.96$ eV/atom) and it comes out to be $0.70$
eV/atom. The calculated $\Delta E_{cohesive}$ of $0.7$0 eV/atom
quantifies the energy required for the formation of considered NV-hg-C$_{3}$N$_{4}$
QD, and it is consistent with the calculated defect formation energy
($0.72$ eV/atom). Following that, the cohesive energies per atom
for all the H-adsorbed configurations are computed using Eq.\ref{eq:2}
and presented in Table \ref{tab:cohesive}. Although, $E_{cohesive}$
is smaller after H adsorption, but its negative sign suggests that
the configurations are thermodynamically stable. 

Dynamic stability is another crucial factor that determines the overall
stability of any system. In order to examine this, vibrational frequencies
are computed for the considered NV-hg-C$_{3}$N$_{4}$ QD, and also,
for all the H-adsorbed configurations. Total $3N-6$ vibrational modes
are possible for each of the configurations, where $N$ denotes the
total number of atoms present in that structure. As a result, $198$
($201$) vibrational frequencies are obtained for the structure(s)
without (with) H-adsorption. The minimum vibrational frequencies presented
in Table \ref{tab:cohesive}, are observed to be real in all the cases.
This implies that our considered NV-hg-C$_{3}$N$_{4}$ QD structure
is dynamically stable as confirmed by the absence of imaginary frequencies,
and its stability is retained under H-adsorbed conditions. The vibrational
properties are studied further by computing their Raman spectra. 

\begin{table}
\caption{\label{tab:cohesive} The calculated cohesive energies per atom, $E_{cohesive}$
and the minimum vibrational frequencies, $Freq$ for the NV-hg-C$_{3}$N$_{4}$
QD, and all the H-adsorbed configurations.}

\centering{}%
\begin{tabular}{ccc}
\toprule 
Configurations & $E_{cohesive}$ (eV) & $Freq$ ($cm^{-1}$)\tabularnewline
\midrule
\midrule 
$NV-hg-C_{3}N_{4}$ & -7.260 & 8.29\tabularnewline
$top-C$ & -7.166 & 8.74\tabularnewline
$top-N$ & -7.166 & 8.71\tabularnewline
$hole$ & -7.167 & 7.76\tabularnewline
$edge-C$ & -7.167 & 7.75\tabularnewline
$edge-N$ & -7.175 & 7.84\tabularnewline
$edge-H$ & -7.175 & 8.70\tabularnewline
$penta-C$ & -7.204 & 5.13\tabularnewline
\bottomrule
\end{tabular}
\end{table}

\subsubsection{Raman Spectra}

\begin{figure}
\begin{centering}
\includegraphics[scale=0.5]{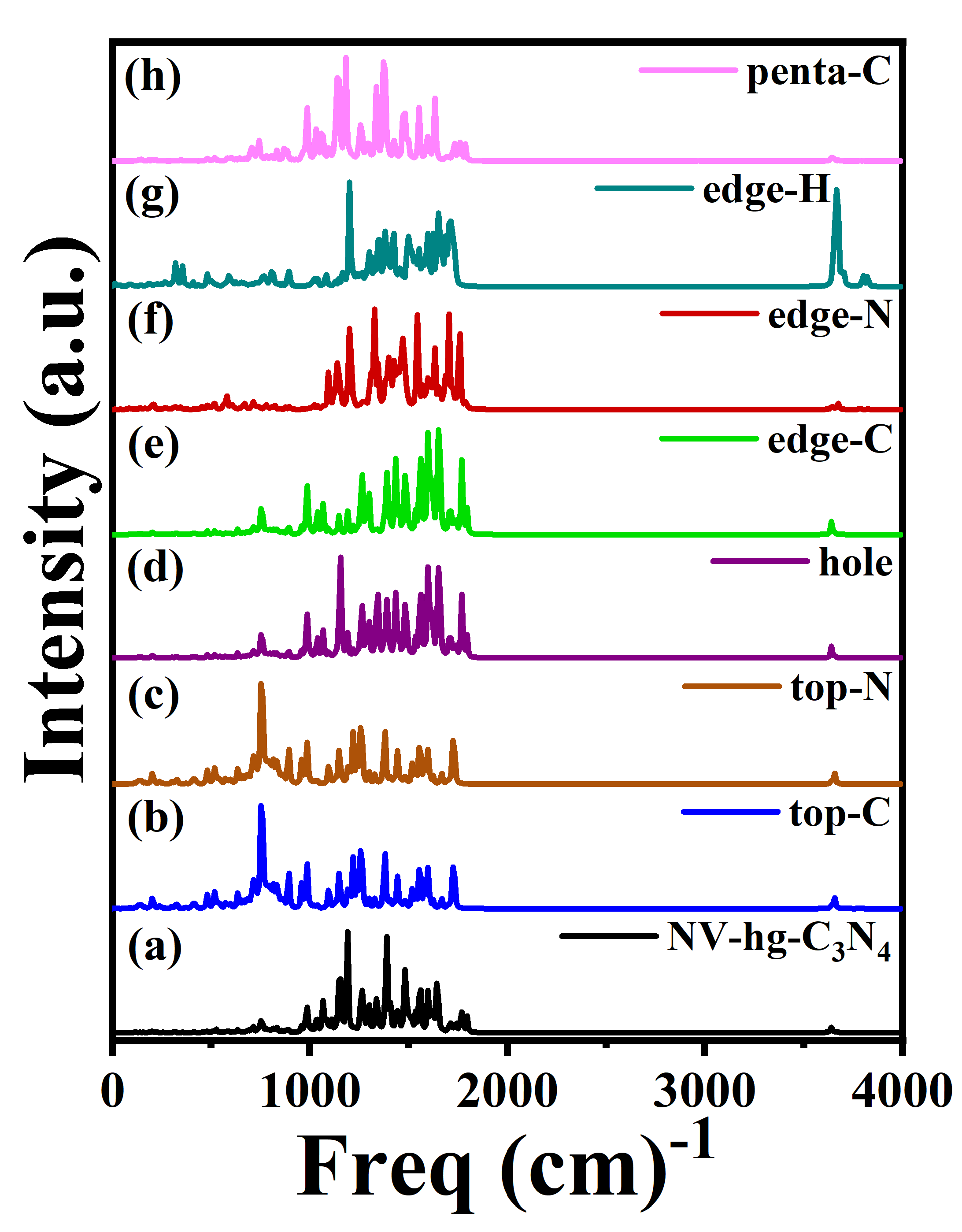}
\par\end{centering}
\caption{\label{fig:3}The calculated Raman Spectra for (a) NV-hg-C$_{3}$N$_{4}$
structure, and (b) $top-C$, (c) $top-N$, (d) $hole$, (e) $edge-C$,
(f) $edge-N$, (g) $edge-H$, and (h) $penta-C$, configurations. }
\end{figure}

The Raman spectra plotted for the considered NV-hg-C$_{3}$N$_{4}$
QD and all the configurations resulting from the adsorption of an
H-atom at different sites are illustrated in Fig.\ref{fig:3}. As
compared to the Raman spectrum of hg-C$_{3}$N$_{4}$ QD (without
defects) which is computed in our previous work \citep{Dange2024},
the number of peaks get enhanced in the case of the defective NV-hg-C$_{3}$N$_{4}$
QD (Fig.\ref{fig:3}(a)). The enhancement of peaks in the nitrogen-vacated
sample compared to the pristine one is also observed experimentally
in the case of NV-hg-C$_{3}$N$_{4}$ nanosheet \citep{LIAO2021}.
In addition, the vibrational peaks are obtained in the range $1000-1800$
$cm^{-1}$, close to the experimentally realized ones ($1200-1700$
$cm^{-1}$) \citep{LIAO2021}. All the peaks within this range, especially
the most intense ones at $1170$ $cm^{-1}$ and $1400$ $cm^{-1}$,
arise due to the stretching modes of $C-N$, $C=N$, and $C-C$ bonds,
the scissoring of $N-C=N,$ $C-N=C$, and $H-N-H$ bonds, and the
rocking of $NH_{2}$ groups present at edges. In a similar manner,
the vibrational modes of the H-adsorbed configurations are analysed
and it is observed that the peaks obtained in all these cases are
also due to the stretching and bending (scissoring and rocking) vibrations
similar to the spectrum before H-adsorption. The prominent peaks are
obtained in $700-1750$ $cm^{-1}$ for the $top-C$ and $top-N$ configurations,
whereas for the $hole$, $edge-C$, $edge-N$, and $penta-C$ configurations,
significant peaks are in $900-1800$ $cm^{-1}$. However, slightly
distinct characteristics are identified for the $edge-H$ configuration.
In addition to the comparatively narrowed frequency region ($1200-1700$
$cm^{-1}$), an intense peak appeared at $3665$ $cm^{-1}$ due to
the stretching of those $N-H$ bonds that link the two heptazine units,
and the nitrogen of such bonds is a part of hexagon (and not of pentagon).

\subsection{Adsorption Energies and Electronic Properties}

After studying the structural and vibrational properties, the adsorption
energies ($\Delta E_{ads}$) are calculated for all the H-adsorbed
configurations, as it is the criterion to theoretically evaluate the
activity of a certain catalyst for HER. The computed $\Delta E_{ads}$
using Eq.\ref{eq:3}, lies in the range $-0.76$ eV to $-3.37$ eV,
as discussed in Table \ref{tab:adsorp}. However, according to the
Sabatier principle \citep{ooka2021}, only the values close to zero
are preferable for HER. The Sabatier principle has been widely employed
as an important criterion or a quantitative predictive tool for the
design of electrocatalysts and their screening for effective catalytic
performance. This principle states that the binding of the reactant
with the catalyst should neither be too strong nor too weak. Applying
the same principle in case of HER implies that, the energy of the
adsorption of hydrogen on the catalytic surface should be close to
zero. As a consequence, out of all the configurations examined, only
four can be considered from the perspective of HER activity, with
the $\Delta E_{ads}$ values of $-0.76$ eV corresponding to the $top-C$
and $top-N$ sites, and $-0.79$ eV corresponding to the $hole$ and
$edge-C$ sites. Therefore, we have investigated the electronic properties,
and most importantly, the HER properties in detail, only for these
four configurations of H-adsorption on the NV-hg-C$_{3}$N$_{4}$
QD. 

\begin{table}
\caption{\label{tab:adsorp}Calculated adsorption energies, $\Delta E_{ads}$
for the H-adsorbed configurations.}

\centering{}%
\begin{tabular}{cc}
\toprule 
Configurations & $\Delta E_{ads}$(eV)\tabularnewline
\midrule
\midrule 
$top-C$ & -0.76\tabularnewline
$top-N$ & -0.76\tabularnewline
$hole$ & -0.79\tabularnewline
$edge-C$ & -0.79\tabularnewline
$edge-N$ & -1.34\tabularnewline
$edge-H$ & -1.34\tabularnewline
$penta-C$ & -3.37\tabularnewline
\bottomrule
\end{tabular}
\end{table}

\begin{figure}[H]
\begin{centering}
\includegraphics[scale=0.4]{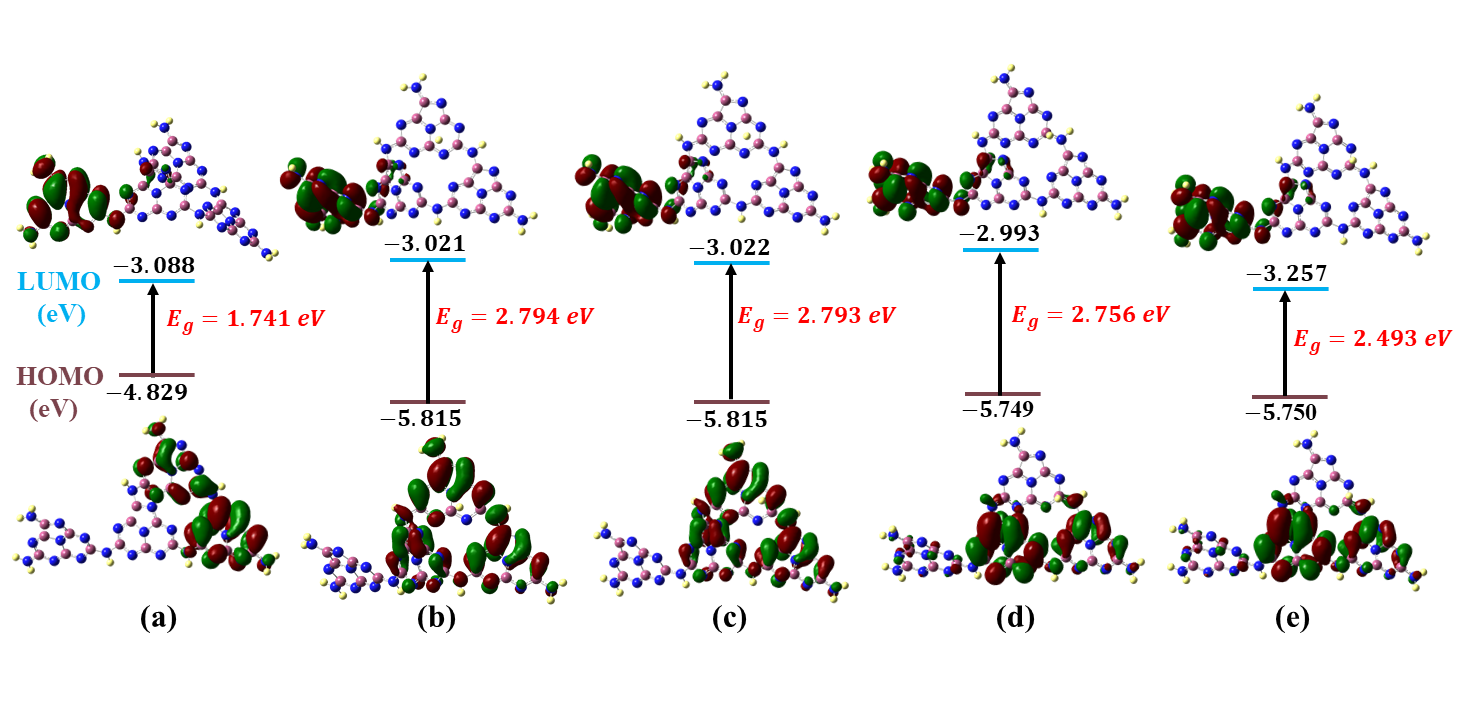}
\par\end{centering}
\caption{\label{fig:4}HOMO and LUMO surfaces for the considered NV-hg-C$_{3}$N$_{4}$
structure (a) before H-adsorption, and after H-adsorption at the sites
(b) $top-C$, (c) $top-N$, (d) $hole$, and (e) $edge-C$. }
\end{figure}

Next, discussing the electronic properties such as energies of the
highest occupied molecular orbital (HOMO) and lowest unoccupied molecular
orbital (LUMO), energy gap ($E_{g}$), work function $(\phi)$, and
charge transfer for the pristine NV-hg-C$_{3}$N$_{4}$ QD and the
four energetically favorable sites for H-adsorption according to the
Sabatier principle, as already mentioned. The energies corresponding
to the HOMO ($E_{HOMO}$) and LUMO ($E_{LUMO}$) along with their
difference, i.e., HOMO-LUMO gap $E_{g}$, are presented in Table \ref{tab:electronic},
while the HOMO and LUMO isosurfaces are plotted in Fig.\ref{fig:4}.
The calculated $E_{g}$ for the NV-hg-C$_{3}$N$_{4}$ QD is $1.741$
eV, whereas before inducing vacancies i.e., for the corresponding
hg-C$_{3}$N$_{4}$ QD, it is $3.8$ eV as reported in our previous
work \citep{Dange2024}. The lowering of $E_{g}$ is due to the lattice
disorder driven by defect formation \citep{LIAO2021}, along with
the change in the electronic structure, causing a shift of the HOMO
and LUMO levels towards each other. The narrowed $E_{g}$ will lead
to enhanced electrical conductivity of the electrons, a prominent
factor controlling the rate of HER. Therefore, the reduced $E_{g}$
due to defect formation, validates our approach of using the ``N-vacated''
hg-C$_{3}$N$_{4}$ QD model and not the one without vacancies for
evaluating their catalytic properties. After H-adsorption, shifting
of the HOMO and LUMO levels is observed. The HOMO (LUMO) level represents
the electron donor (acceptor) properties of the system or the nature
of the nucleophilic (electrophilic) kind. The visualization of the
HOMO and LUMO isosurfaces presented in Fig.\ref{fig:4} demonstrates
that both the HOMO and LUMO are delocalized over multiple $C-N$ bonds
in all the studied configurations. It is also observed that there
is no significant electronic density near the adsorbed hydrogen atom
in the case of all the four sites, indicating the transfer of charge
from this hydrogen atom to the rest of the structure. This is also
confirmed by the calculated charge transfer between the NV-hg-C$_{3}$N$_{4}$
QD moiety and the adsorbed hydrogen atom, presented in Table \ref{tab:electronic}.
The HOMO level shifts downward due to H-adsorption at all the four
sites, indicating reduced electron-donor ability, whereas, the LUMO
level shifts downward (upward) for the $edge-C$ (other adsorption
sites), indicating enhanced (reduced) electron-acceptor ability. Also,
the shifts are uneven which results in the modification of $E_{g}$.
In the case of lower $E_{g}$, electrons can be readily promoted to
the LUMO level which makes them more mobile leading to efficient participation
in the reduction of water molecules to hydrogen gas via any of the
two possible reaction mechanism, namely, Volmer-Heyrovsky and Volmer-Tafel,
discussed later in detail. After adsorption of hydrogen atom at all
the considered energetically favorable sites, $E_{g}$ increases and
lies within $2.498$ eV -- $2.794$ eV. However, these energy gaps
are smaller than that of the pristine hg-C$_{3}$N$_{4}$ QD ($3.8$
eV), and thus the electrons retain relatively good conductivity even
after H-adsorption on the NV-hg-C$_{3}$N$_{4}$ QD. The work function
is a crucial parameter that defines the energy needed to remove an
electron from a solid and transport it to a point just outside the
surface, typically in the vacuum. Thus, to get a deeper insight into
the stability of the considered system after H-adsorption, the work
function $(\phi)$ has been calculated using the formula:

\begin{equation}
\phi=\frac{IP+EA}{2}
\end{equation}

where $IP=-E_{HOMO}$ and $EA=-E_{LUMO}$ are the estimated ionization
potential and electron affinity of the system, respectively. The obtained
values of $\phi$ are reported in Table \ref{tab:electronic}, from
which it is observed that the $\phi$ gets increased after H-adsorption
and lies within $4.418$ eV -- $4.503$ eV. Higher $\phi$ suggests
higher stability of the structure after H-adsorption. Furthermore,
the obtained $\phi$ values are comparable with those of Pt and Pd-based
catalysts, implying that our proposed hg-C$_{3}$N$_{4}$ QD could
be a viable option for efficient HER performance. 

\begin{table}
\caption{\label{tab:electronic}Calculated energies of HOMO ($E_{HOMO}$) and
LUMO ($E_{LUMO}$), energy gap ($E_{g}$), charge transfer $(Q)$,
and work function ($\phi$) for the studied configurations.}

\centering{}%
\begin{tabular}{cccccc}
\toprule 
Configurations & $E_{HOMO}$(eV) & $E_{LUMO}$(eV) & $E_{g}$(eV) & $Q$ (e) & $\phi$(eV)\tabularnewline
\midrule
\midrule 
$NV-hg-C_{3}N_{4}$ & -4.829 & -3.088 & 1.741 & --- & 3.958\tabularnewline
$top-C$ & -5.815 & -3.021 & 2.794 & -0.213 & 4.418\tabularnewline
$top-N$ & -5.815 & -3.022 & 2.793 & -0.354 & 4.418\tabularnewline
$hole$ & -5.749 & -2.993 & 2.756 & -0.035 & 4.371\tabularnewline
$edge-C$ & -5.750 & -3.257 & 2.498 & -0.006 & 4.503\tabularnewline
\bottomrule
\end{tabular}
\end{table}

\begin{figure}[H]
\begin{centering}
\includegraphics[scale=0.4]{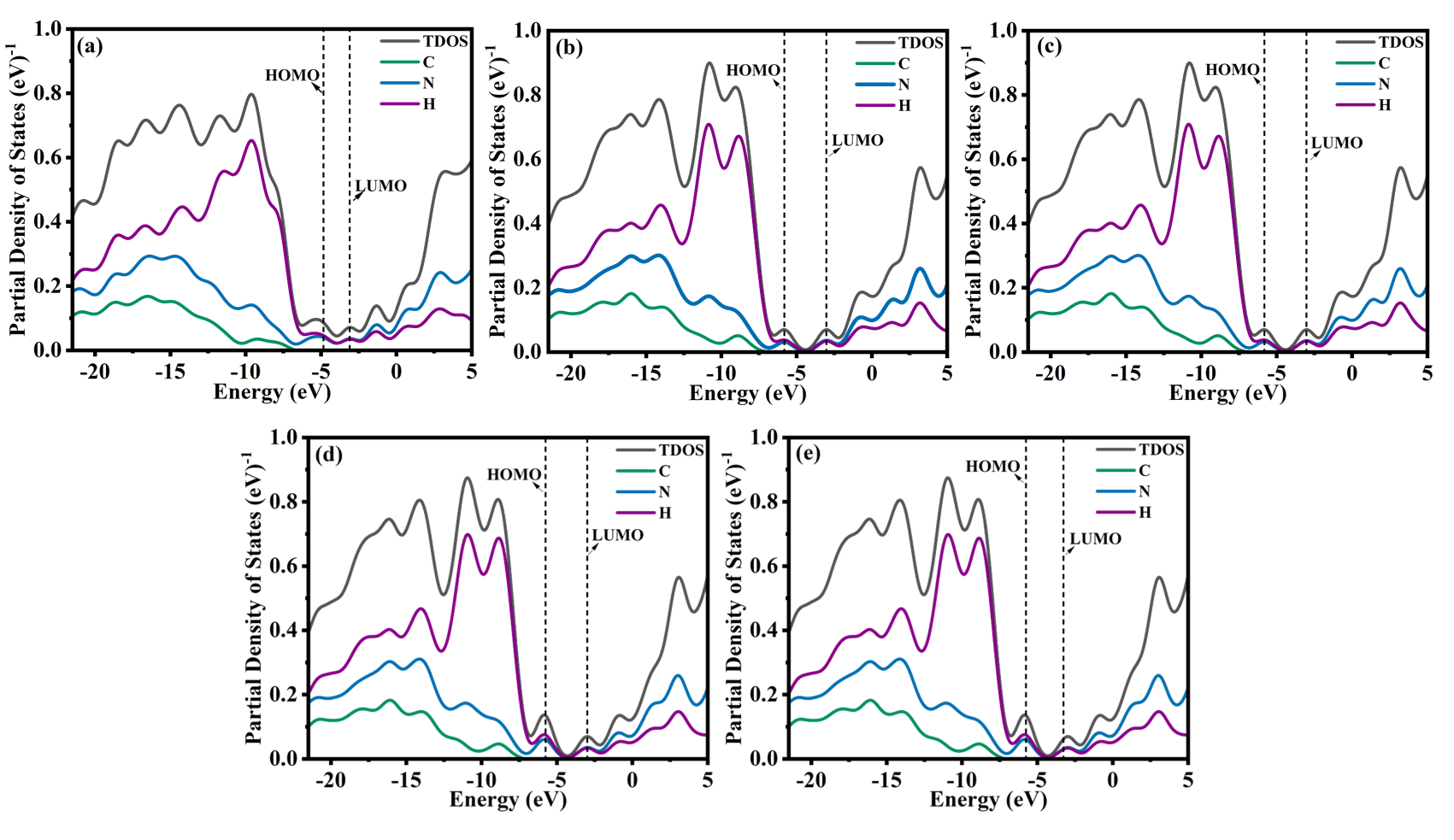}
\par\end{centering}
\caption{\label{fig:5} Total density of states (TDOS) and partial density
of states (PDOS) plotted for NV-hg-C$_{3}$N$_{4}$ QD (a) before
H-adsorption, and after H adsorption at the sites (b) $top-C$, (c)
$top-N$, (d) $hole$, and (e) $edge-C$.}
\end{figure}

Fig.\ref{fig:5} presents the total density of states (TDOS) and partial
density of states (PDOS) for the considered structure before and after
H-adsorption at different sites. While the TDOS provides information
on the occupied and unoccupied electronic energy levels, the PDOS
separately indicates the contribution of each constituent element
to the TDOS. The number of peaks in TDOS get reduced in the HOMO region
after H-adsorption, whereas no significant change is observed in the
LUMO region. From the PDOS, it is clear that the contribution of carbon,
nitrogen and hydrogen atoms in the HOMO region are in the order, H
> N > C. However, there is distinct behavior observed in the LUMO
region, where nitrogen atoms make the dominant contribution, followed
by hydrogen atoms. Additionally, it is noteworthy that carbon atoms
do not contribute in the LUMO region, which is similar to the case
of pristine hg-C$_{3}$N$_{4}$ QD \citep{Dange2024}. In addition,
no significant change in the qualitative nature of TDOS and PDOS plots
of NV-hg-C$_{3}$N$_{4}$ QD is observed after H-adsorption at all
the four sites. Therefore, it implies that the electronic structure
of NV-hg-C$_{3}$N$_{4}$ QD remains almost unaltered after H-adsorption. 

\subsection{Hydrogen Evolution Reaction}

Prior to discussing the HER-related properties of the considered structures,the
mechanism of electrocatalytic water splitting for HER is discussed
concisely in the following. The splitting of water necessitates a
certain level of overpotential, which must be provided by means of
the two electrodes, anode and cathode. The HER which involves the
reduction of water at the cathode to produce hydrogen, is a multiphase
reaction process. It can proceed via the two possible reaction mechanisms:
(a) Volmer-Heyrovsky and (b) Volmer-Tafel, either in an acidic or
an alkaline environment. These two mechanisms result from the three
different possible reaction steps. In an acidic electrolyte, the first
mandatory step, i.e., the Volmer reaction: $H^{+}+e^{-}\rightarrow H_{ads}$
takes place when an electron on the catalytic (cathode) surface captures
a proton from electrolyte to form an intermediate state of the adsorbed
hydrogen atom ($H_{ads}$). After this step, the evolution of hydrogen
can occur via the Heyrovsky step ($H_{ads}+H^{+}+e^{-}\rightarrow H_{2}$)
or the Tafel step ($2H_{ads}\rightarrow H_{2}$). 

\begin{figure}[H]
\begin{centering}
\includegraphics[scale=0.6]{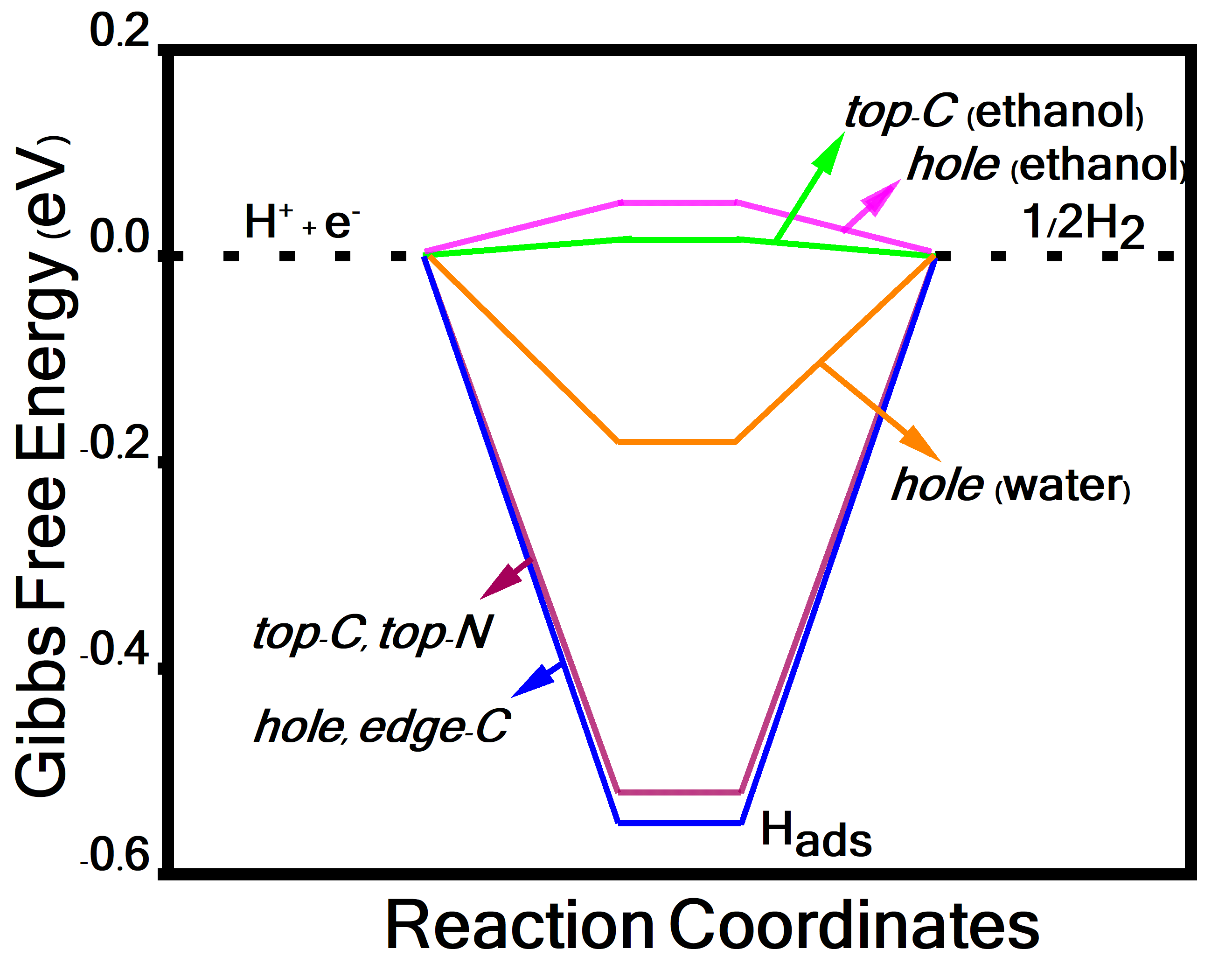}
\par\end{centering}
\caption{\label{fig:6}Free Energy diagram for HER corresponding to all the
favorable sites of H-adsorption ($H_{ads}$). }

\end{figure}

Thus, the entire HER process involves adsorption of hydrogen, $\Delta E_{ads}$
and the corresponding Gibbs free energy ($\Delta G$) plays an important
role in deciding the catalytic activity of any material. The favorable
$\Delta E_{ads}$ according to the Sabatier principle \citep{ooka2021}
has already been discussed in the previous section. Following that,
the optimum $\Delta G$ value for efficient HER performance should
approach zero according to the Sabatier principle, i.e., $\Delta G\approx$0.
If $\Delta G<<0$, hydrogen atom gets adsorbed too strongly so that
it would be difficult to desorb it as needed for further reaction,
and if $\Delta G>>0$, no adsorption of hydrogen would take place.
$\Delta G$ is calculated for the four energetically favorable configurations
of H-adsorption on the considered NV-hg-C$_{3}$N$_{4}$ QD using
Eq.\ref{eq:4}, and its values are listed in Table \ref{tab:Calculated-Gibb's-free}.
The $\Delta G$ value is $-0.52$ eV when H-atom is placed over the
top of a carbon and a nitrogen atom, while it is $-0.55$ eV when
placed over the $hole$ position, and over a carbon atom present at
an edge. Fig.\ref{fig:6} illustrates the free energy diagram for
HER, starting from the initial $H^{+}+e^{-}$ resulting in evolved
hydrogen via intermediate adsorbed H-atom ($H_{ads}$). Afterwards,
the key quantity, namely the overpotential ($\eta$) is calculated
using Eq.\ref{eq:5}. $\eta$ is defined as the minimum overpotential
required to facilitate all the reaction steps involved in HER. As
mentioned earlier, only absolute values of $\eta$ are considered,
which implies that a lower value of $\eta$ indicates higher catalytic
performance. The two smallest values of $\eta$ ($0.52$ V and $0.55$
V) are obtained for four different H-adsorbed sites as presented in
Table \ref{tab:Calculated-Gibb's-free}. These results suggest that
the nitrogen-vacated $hg-C_{3}N_{4}$ QDs can be used as a moderately
effective electrocatalyst for HER performance. 

\begin{table}
\caption{\label{tab:Calculated-Gibb's-free}Calculated Gibbs free energy ($\Delta G$),
overpotential ($\eta)$ and open circuit voltage ($V$) for the considered
H-adsorbed configurations which are energetically favorable.}

\begin{centering}
\begin{tabular}{cccc}
\toprule 
Configurations & $\Delta G$(eV) & $\eta$(V) & $V$(V)\tabularnewline
\midrule
\midrule 
$top-C$ & -0.52 & 0.52 & -0.02\tabularnewline
$top-N$ & -0.52 & 0.52 & -0.02\tabularnewline
$hole$ & -0.55 & 0.55 & -0.07\tabularnewline
$edge-C$ & -0.55 & 0.55 & -0.06\tabularnewline
\bottomrule
\end{tabular} 
\par\end{centering}
\end{table}

The open circuit voltage ($V$) has also been calculated employing
the standard hydrogen electrode (SHE) as a reference, using the formula

\begin{equation}
V=\frac{\phi}{e}-4.44,
\end{equation}

where $\phi$ represents the work function calculated above, and $4.44$
is the absolute potential of SHE \citep{Trasatti1986}. The obtained
values of $V$ lie in the range $-0.02$ V to $-0.07$ V for the considered
H-adsorbed configurations, as tabulated in Table \ref{tab:Calculated-Gibb's-free}.

\subsubsection{HER in the presence of solvents}

After calculating the HER-related properties for the structure under
consideration by treating it in vacuum, our interest is to study the
impact of solvents on the said properties. Formation of hg-C$_{3}$N$_{4}$
QDs from hg-C$_{3}$N$_{4}$ nanosheet results in the introduction
of oxygenated functional groups, such as carboxylate groups at the
edges, as confirmed by the strong peak at $1385$ $cm^{-1}$in FT-IR
spectrum \citep{wang2014}. These hydrophilic functional groups render
hg-C$_{3}$N$_{4}$ QDs water-soluble \citep{wang2014}, and the same
concept can be applied to predict their solubility in ethanol. Additionally,
studies have reported the use of ethanol during the synthesis of g-C$_{3}$N$_{4}$
nanoarchitectures \citep{JIANG2019,YANG2022}. Motivated by these
findings, we aim to investigate the HER performance of the NV-hg-C$_{3}$N$_{4}$
QD under consideration by treating them in ethanol. 

\begin{figure}[H]
\begin{centering}
\includegraphics[scale=0.4]{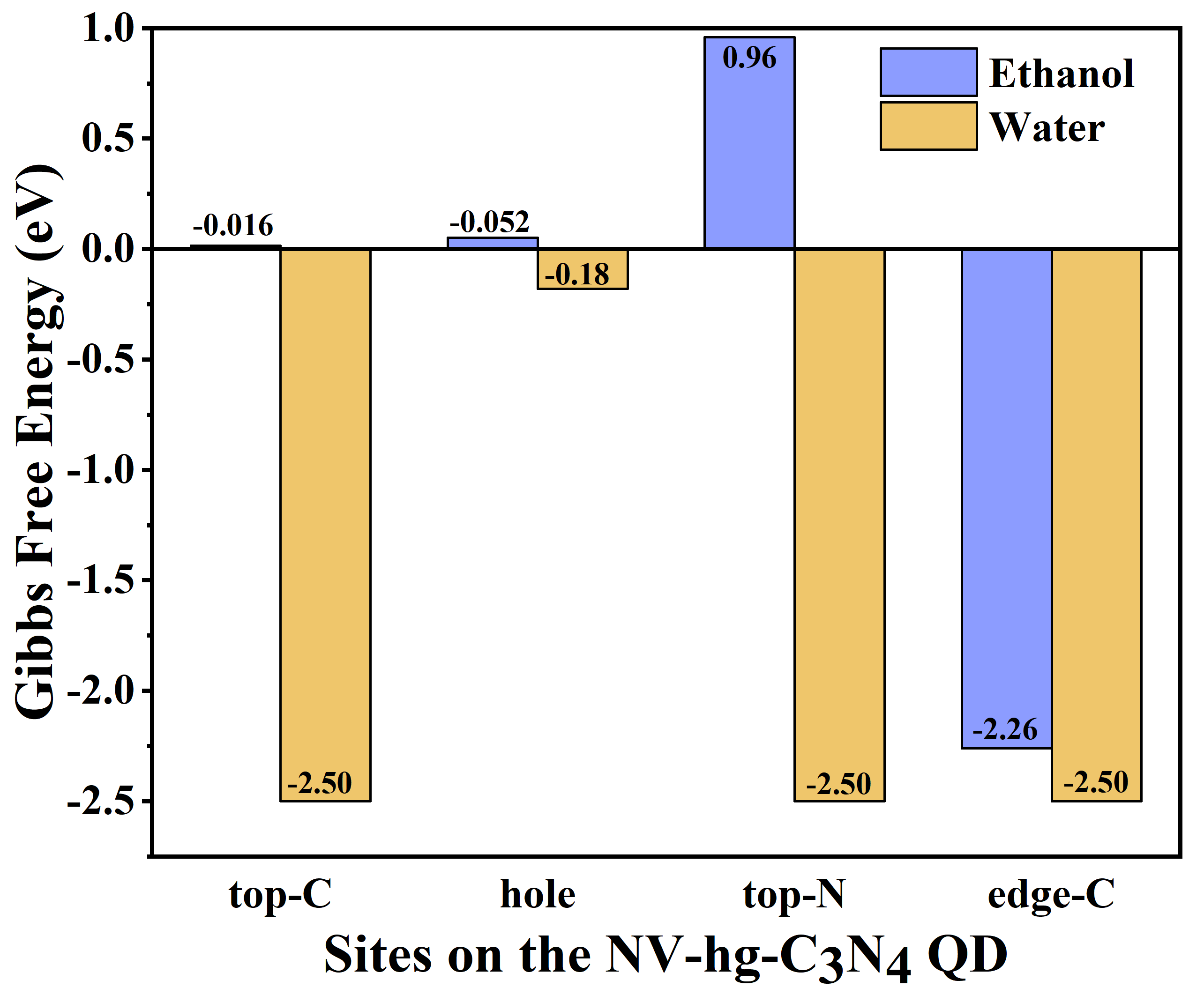}
\par\end{centering}
\caption{\label{fig:solvent-graph}Gibbs free energies ($\Delta G$) of the
four H-adsorbed configurations when dissolved in ethanol and water. }
\end{figure}

To investigate this, we conducted geometry optimization on the four
selected configurations of H-adsorption, out of the seven sites considered
initially. This optimization has been performed in the presence of
ethanol using the IEFPCM model. The obtained $\Delta E_{ads}$ values
are presented in Table \ref{tab:solvent-table}. The results indicate
that the extent of adsorption at the $edge-C$ site is quite strong
so it would not be favorable for HER. Further, a significant positive
$\Delta E_{ads}$ in the case of site $top-N$, also indicates its
low activity. Therefore, only two sites, i.e., $top-C$ and $hole$,
can be considered as active sites for H-adsorption on the NV-hg-C$_{3}$N$_{4}$
QD in the presence of ethanol. As a consequence, $\Delta G$ ($\eta$)
is calculated for these two active sites and the obtained values are
$0.016$ eV ($16$ mV) and $0.052$ eV ($52$ mV) for the $top-C$
and $hole$ configurations, respectively. The $\Delta G$ for these
two cases is also depicted in Table \ref{tab:solvent-table} and Fig.\ref{fig:solvent-graph}
along with that for other two sites, for comparison. The free energy
diagram for the two active sites, namely, $top-C$ and $hole$ in
the presence of ethanol is included in Fig.\ref{fig:6}. Although
the number of active sites of the considered $NV-hg-C_{3}N_{4}$ QD
get reduced when treated in ethanol, the free energy and the corresponding
overpotential also get reduced significantly for the two favorable
sites. The significant reduction in the overpotential due to the change
in dielectric environment of the considered $NV-hg-C_{3}N_{4}$ QD
by treating it in ethanol can be explained on the basis of charge
transfer. The amount of charge transfer ($Q$) between the adsorbed
H-atom and the considered $NV-hg-C_{3}N_{4}$ QD moiety in the case
of vacuum is already discussed (see Table \ref{tab:electronic}).
As discussed next, the inclusion of ethanol as a solvent resulted
in a significant change in the amount of charge transfer with respect
to the QD in vacuum, denoted by $\Delta Q=Q_{ethanol}-Q_{vacuum}$,
where $Q_{ethanol}$ ($Q_{vacuum}$) is the charge transfer in case
of ethanol (vacuum). It is observed that, $\Delta Q_{top-C}=-0.18e$
and $\Delta Q_{hole}=-0.03e$, implying that: (a) the Coulomb interactions
between the adsorbed H-atom and the considered catalyst, and therefore,
$\Delta G$ and $\eta$, will be reduced for these two sites when
dissolved in ethanol as compared to vacuum, and (b) the amount of
reduction for the $top-C$ site will be much more as compared to the
$hole$ site leading to $\eta_{top-C}(16\,mV)<\eta_{hole}(52\,mV)$.
For the remaining two sites, $top-N$ and $edge-C$, opposite is seen,
that is charge transfer increases with $\Delta Q>0$, resulting in
stronger Coulomb interactions, thereby, increased values of $\Delta G$
and $\eta$. This is fully consistent with our numerical results with
$\Delta Q_{top-N}(0.54e)<\Delta Q_{edge-C}(0.67e)$, resulting in
$\eta_{top-N}(0.96\,V)<\eta_{edge-C}(2.26\,V)$. The obtained $\eta$
is very close to zero in the presence of ethanol, with the minimum
value of $16$ mV for the $top-C$ configuration. Such a low value
of overpotential implies that the cost-effective, abundant NV-hg-C$_{3}$N$_{4}$
QDs have the potential to replace traditional metal-based electrocatalysts
for HER. 

Next, we have also considered water as a solvent, and the reactivity
of the considered NV-hg-C$_{3}$N$_{4}$ QD immersed in water is checked
for HER. The calculations are performed on the four configurations
of H-adsorption which we considered in the case of ethanol as well.
The obtained $\Delta E_{ads}$ values are reported in Table \ref{tab:solvent-table},
from which we infer that when placed at the $top-C$, $top-N$, and
$edge-C$ sites, H-atom gets adsorbed much strongly when dissolved
in water, compared to the other two cases, i.e., ethanol and vacuum.
The strong adsorption is due to the bond formation of the adsorbed
H-atom with a neighboring C-atom in all the three cases, thus giving
same $\Delta E_{ads}$ (-2.74 eV). The corresponding $\Delta G$ (-2.50
eV) and $\eta$ (2.50 V) values are quite large and significantly
higher than those in the cases of QD in vacuum as well as when dissolved
in ethanol (see Table \ref{tab:solvent-table} and Table \ref{tab:Calculated-Gibb's-free}).
Thus these three sites ($top-C$, $top-N$, and $edge-C$) of the
considered NV-hg-C$_{3}$N$_{4}$ QD are not suitable for HER activity.
In the case of $hole$ site, $\Delta E_{ads}$ is less negative (-0.42
eV) and close to zero when treated in water compared to that in vacuum.
The corresponding $\Delta G$ (-0.18 eV) and $\eta$ (0.18 V) values
suggest that the $hole$ site will be effective for HER activity when
the considered NV-hg-C$_{3}$N$_{4}$ QD is immersed in water, and
thus it is included in the free energy diagram of Fig.\ref{fig:6}.
Further, it is observed that the use of water as a solvent gives only
one active site, i.e., $hole$ for HER and that too is less active
($\eta$ = 0.18 V) compared to that in the case of ethanol ($\eta$
= 52 mV). The order of HER activity of the considered catalyst corresponding
to the $hole$ site in different dielectric environments is as follows:
ethanol > water > vacuum. Therefore, we infer that out of the two
solvents considered in this work, ethanol is much superior in increasing
the electrocatalytic activity of the NV-hg-C$_{3}$N$_{4}$ QD under
consideration.

\begin{table}[H]
\caption{\label{tab:solvent-table}Calculated adsorption energies ($\Delta E_{ads}$),
Gibbs free energies ($\Delta G$), and overpotential ($\eta$) for
the H-adsorbed configurations of the NV-hg-C$_{3}$N$_{4}$ QD treated
in ethanol and water.}

\centering{}%
\begin{tabular}{ccccccc}
\toprule 
\multirow{2}{*}{Configurations} & \multicolumn{3}{c}{Ethanol} & \multicolumn{3}{c}{Water}\tabularnewline
\cmidrule{2-7} \cmidrule{3-7} \cmidrule{4-7} \cmidrule{5-7} \cmidrule{6-7} \cmidrule{7-7} 
 & $\Delta E_{ads}$ (eV) & $\Delta G$ (eV) & $\eta$ (V) & $\Delta E_{ads}$ (eV) & $\Delta G$ (eV) & $\eta$ (V)\tabularnewline
\midrule
\midrule 
$top-C$ & -0.22 & 0.016 & 0.016 & -2.74 & -2.50 & 2.50\tabularnewline
$top-N$ & 0.72 & 0.960 & 0.960 & -2.74 & -2.50 & 2.50\tabularnewline
$hole$ & -0.18 & 0.052 & 0.052 & -0.42 & -0.18 & 0.18\tabularnewline
$edge-C$ & -2.50 & -2.260 & 2.260 & -2.74 & -2.50 & 2.50\tabularnewline
\bottomrule
\end{tabular}
\end{table}

\section{CONCLUSION}

To summarize, we investigated nitrogen-vacancies induced hg-C$_{3}$N$_{4}$
QDs as an electrocatalyst for hydrogen evolution during water splitting.
The hg-C$_{3}$N$_{4}$ QD made of the four heptazine units has been
considered along with some nitrogen-vacancies (NV-hg-C$_{3}$N$_{4}$
QD), to reflect a realistic scenario. Out of the seven possible H-adsorption
sites, four exhibited favorable energetics, leading to the calculation
of their electronic and HER-related properties. As far as the electronic
properties are concerned, HOMO and LUMO energy levels, energy gap,
charge transfer, work function, and density of states are studied.
The $\Delta G$ and $\eta$ are computed for the four configurations
of H-adsorption. The results showed that the NV-hg-C$_{3}$N$_{4}$
QD is a promising electrocatalyst for HER, with moderately low values
of $\eta$ ($-0.52$ V for $top-C$ and $top-N$, and $-0.55$ V for
$hole$ and $edge-C$ configurations). The effect of ethanol and water
as solvents on the catalytic properties of the NV-hg-C$_{3}$N$_{4}$
QD is also investigated. In comparison to the QD in vacuum, when dissolved
in water, only one active site, $hole$, exhibits a reduction in overpotential
from 0.55 V to 0.18 V. Whereas, in the case of ethanol, two active
sites, $top-C$ and $hole$ have much smaller overpotentials of $16$
mV and 52 mV, respectively. These findings suggest that N-deficient
hg-C$_{3}$N$_{4}$ QDs dissolved in ethanol can be excellent candidates
for HER performance, and our quantitative results can guide the experimental
efforts in its practical realization as an economically feasible electrocatalyst.
\begin{suppinfo}
The Supporting Information is available free of charge on the ACS
Publications website at DOI:

The given supporting information contains two figures: (a) Nitrogen-vacancy
formation at three different possible sites of a single unit of the
heptazine structure, and (b) Seven different sites of the NV-hg-C$_{3}$N$_{4}$
QD considered for the adsorption of H-atom. It also contain a brief
discussion of the HER parameters investigated initially for the pristine
hg-C$_{3}$N$_{4}$ QD (without any defects).
\end{suppinfo}

\section*{Author Information }

\subsection*{Corresponding Authors}

Alok Shukla:  {*}E-mail: shukla@phy.iitb.ac.in

\subsection*{Author Contributions}

The manuscript was written through contributions of all authors. All
authors have given approval to the final version of the manuscript.

\section*{Notes}

The authors declare no competing financial interest.
\begin{acknowledgement}
One of the authors, K.D. acknowledges financial assistance from the
Prime Minister Research Fellowship of India (PMRF award ID-1302054).
V.R. acknowledges the support through the Institute Post-Doctoral
Fellowship (IPDF) of Indian Institute of Technology Bombay. 
\end{acknowledgement}
\bibliographystyle{achemso}
\bibliography{Ref}

\pagebreak{}

\includepdf{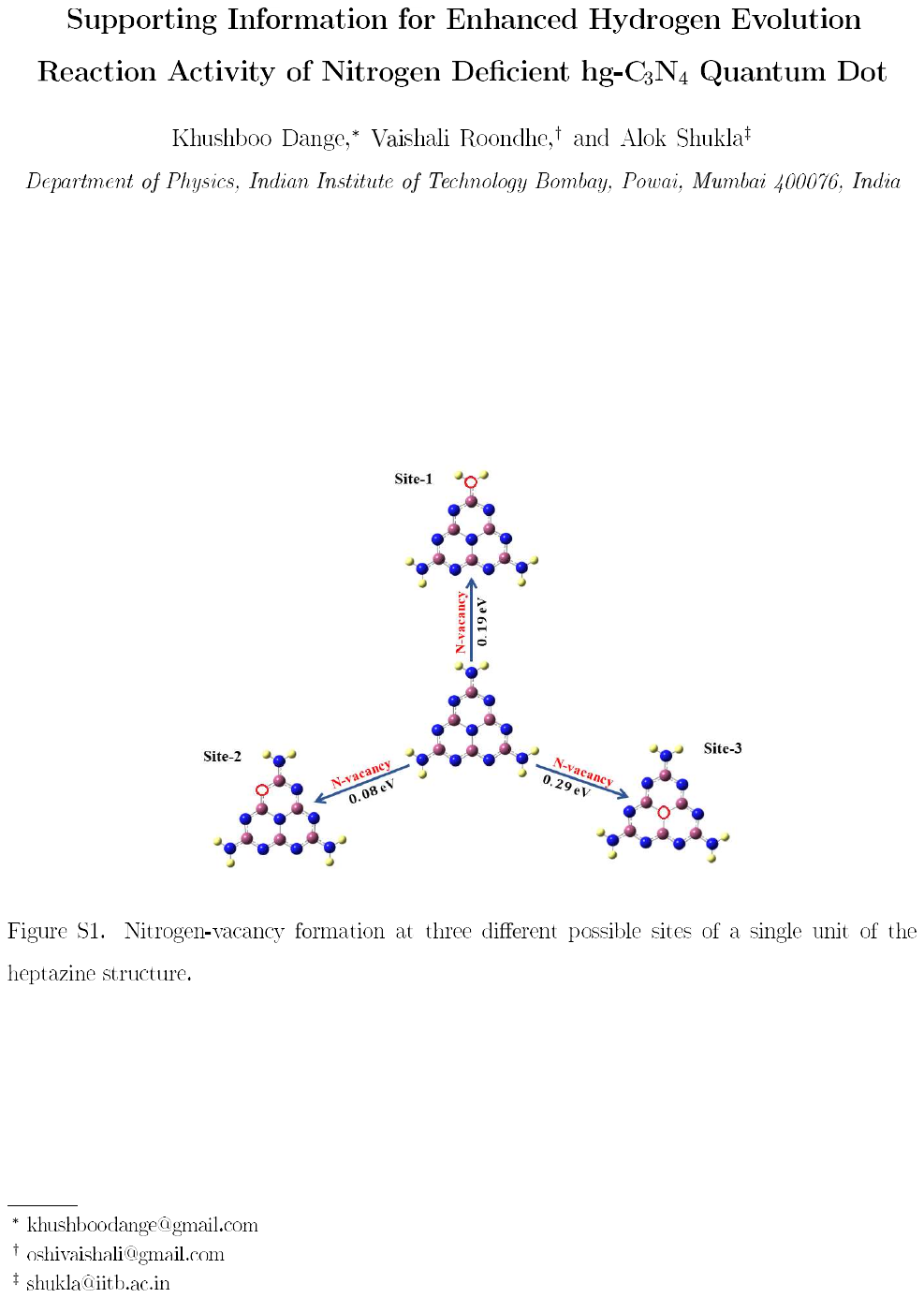}

\includepdf{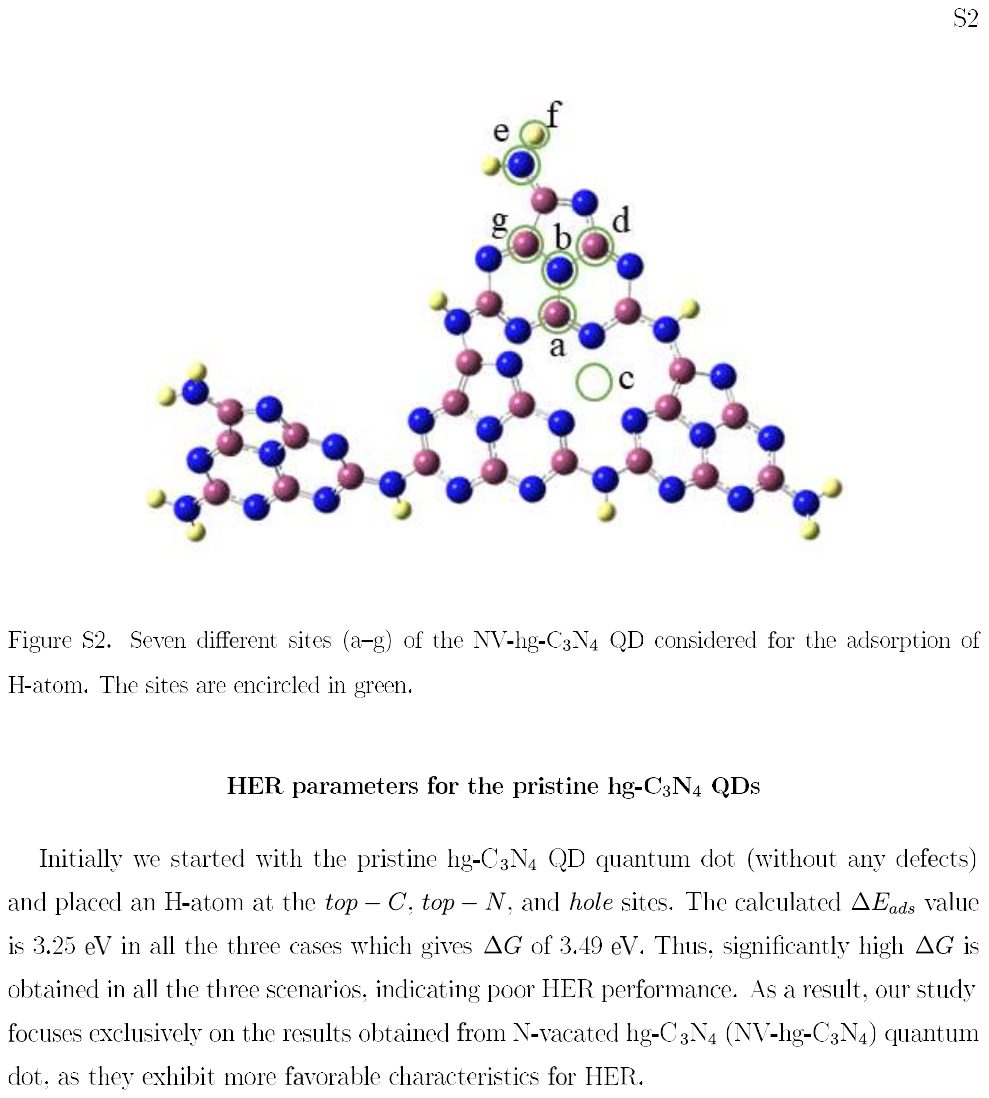}
\end{document}